\documentstyle[aps,eqsecnum,preprint,prbbib]{revtex}
\begin{document}
\draft
\title{Time Scales for Transitions between Free Energy Minima of a Dense
Hard Sphere System}
\author{Chandan Dasgupta}
\address{Department of Physics, Indian Institute of Science \\
Bangalore 560012, India}
\author{Oriol T. Valls}
\address{School of Physics and Astronomy \\
and Minnesota Supercomputer Institute \\
University of Minnesota, Minneapolis, Minnesota 55455}
\date{\today}
\maketitle
\begin{abstract}
Time scales associated with activated transitions between  glassy 
metastable states of a free energy functional appropriate for a dense hard
sphere system are calculated by using a new Monte Carlo method for the 
local density variables.
In particular,
we calculate the time the sytem, initially placed in a shallow glassy minimum
of the free energy, spends in the neighborhood of this minimum before
making a transition to the basin of attraction of another free energy minimum.
This time scale is found to increase as the average density is
increased. We find a crossover density near which this time scale increases
very sharply and becomes longer than the longest times accessible in our
simulation. This time scale 
does not show any evidence 
of increasing with sample size.
\end{abstract}

\section {Introduction}
Although the behavior of supercooled liquids near the 
glass transition has been
studied extensively over several decades \cite{jack,angell}, a complete 
understanding of some of the observed phenomena is not yet available. Existing
theories of the glass transition may be broadly classified into two
categories. The first category consists of theories which describe the glass
transition as a purely dynamic phenomenon. Mode coupling (MC) theories of the
glass transition \cite{mct} are the most prominent ones in this class. In
MC theories, the slowing down of the dynamics near the glass transition is 
attributed to a nonlinear feedback mechanism arising from correlations of
density fluctuations in the liquid. MC theories provide a detailed and
qualitatively correct description of the dynamic behavior observed in 
experiments \cite{exp} and numerical simulations \cite{num} over a temperature
range that covers the first few decades of the growth of the characteristic
relaxation time of so-called ``fragile'' \cite{angell} liquids in the
supercooled regime. The original version of MC theories \cite {leu,bgs} 
predicts a
power-law divergence of the relaxation time at an {\it ideal glass transition
temperature} $T_c$. Experimentally, however, this divergence is not found, and
the predictions of MC theories do not provide a correct description of the 
actual behavior at temperatures close to or lower than the $T_c$ extracted
from power-law fits to the data at higher temperatures. It is generally 
believed that the breakdown of conventional MC theories at temperatures near
$T_c$ arises because ``activated processes'' become important at
such temperatures. The conventional MC formalism has been generalized
\cite{gs,my} to incorporate in a phenomenological way some of the effects
of activated hopping processes, and results of recent light scattering
experiments \cite{cummins} have been interpreted in terms of this extended
version of MC theory. However, the microscopic nature of these activated
processes has not been fully elucidated so far.

In the second class of theories \cite{pgw} 
of the glass transition, the starting point is
the assumption that the 
free energy of the liquid, expressed as a functional of the
local average density, develops a large number of ``glassy'' local mimima as
the temperature is decreased below the equilibrium crystallization temperature.
These glassy minima, characterized by an inhomogeneous but aperiodic 
distribution of the local density, are distinct from the (metastable) liquid
minimum where the average density is uniform and from the (globally stable) 
crystalline minimum where
the average density is a periodic function of position. During a 
quench from the liquid state, the system  gets trapped in one of these
glassy minima from which it eventually relaxes. The relaxation is
slow because it involves
thermally activated transitions over free energy barriers. In this picture,
therefore, the slow dynamics near the glass transition is attributed to 
activated transitions among metastable glassy minima of the free energy. This
description is similar to that developed in recent years for a number
of quenched random systems such as spin glasses \cite{SG}. 
Several authors \cite{sp,ktw,sethna,ssh} have taken
this analogy further and suggested that the behavior observed near the glass
transition is the precursor of a true thermodynamic phase transition which 
would take place at a temperature  lower than the conventional glass transition
temperature $T_g$ (defined \cite{angell} 
as the temperature at which the viscosity reaches
a value of 10$^{13}$ P) if one could maintain thermodynamic equilibrium
all the way down to this temperature. This scenario, however, remains
speculative because the existence of such a phase transition has not yet been
demonstrated convincingly for any physically realistic system.

Recently, we have carried out a number of investigations \cite{us1,us2,us3,us4}
which suggest that 
elements of these two apparently dissimilar descriptions of the glass transition
should be combined for the development of
a full understanding of the observed phenomena. This conclusion was based on
the results of a numerical study of a set of Langevin equations which describe
the nonlinear fluctuating hydrodynamics of a dense hard sphere liquid. 
Information about the static structure of the liquid was incorporated in the
Langevin equations considered by us through a free energy functional of a form
suggested by Ramakrishnan and Yussouff \cite{RY}. From previous studies 
\cite{cd1,cd2}, it is known that this free energy functional develops a large
number of glassy local minima at densities higher than the equilibrium 
crystallization density. (The control parameter for a hard sphere 
system is the dimensionless density $n^* \equiv \rho_0 \sigma^3$, where 
$\rho_0$ is the average
number density and $\sigma$ is the hard-sphere diameter; increasing (decreasing)
$n^*$ has the same effect as decreasing (increasing) the temperature of systems
for which the temperature is the relevant control parameter). Our numerical
study of the dynamics showed \cite{us1,us2} that for relatively small values
of $n^*$ ($n^* < 0.95$),
a system initially prepared in the liquid state remains in the 
vicinity of the metastable liquid minimum during the time scales accessible in
the simulation. The dynamic behavior observed in this regime was found to be
in good qualitative agreement with the predictions of MC theories. At higher
densities (specifically, for $n^* $ greater than a ``crossover density'' $n^*_x
\approx 0.95$), we found\cite{us3} a qualitatively different behavior.
At these densities,  after spending an
initial period of time (which decreases as the density is increased) near the
liquid minimum, the system makes a transition to one of the glassy minima of 
the free energy. This observation implies that the long-time dynamic behavior 
for $n^* > n^*_x$
is not governed by small
fluctuations near the uniform liquid minimum: the
glassy local minima of the free energy have to be taken into account in a
proper description of the dynamics. In particular, activated transitions among
these glassy minima are expected to play a crucial role in the dynamic behavior
in this regime. Although a different
Langevin simulation method could in principle be
developped to study the dynamics in this density range, the methods of Refs.
(\onlinecite{us1,us2,us3,us4}), which are based on a liquid-like formulation,
are not physically appropriate for such a study. In practice,
attempts to use these methods at higher densities lead to numerical
instabilities associated with
large-scale density fluctuations present 
at the inhomogeneous minima. 
For this reason, our Langevin dynamics cannot be carried all the way to
equilibration at densities higher than $n^*_x$, although the early and
intermediate stages of the system's evolution can still be studied\cite
{us3}. Therefore, we develop here an alternative method to study the 
dynamics of the same system at higher densities.

In this paper, we present the results of a study in which we
develop a novel Monte Carlo
method and use it to determine the characteristic time scales of transitions
between glassy free energy minima for $n^* \ge n^*_x$. The Monte Carlo method,
described in detail in the following section, is much simpler to implement
than the Langevin simulation method.  The numerical instabilities
mentioned above are not present in the Monte Carlo procedure used by us. This
simplification, however, is achieved at the cost of abandoning the physical
(coarse grained) dynamics embodied in the Langevin equations (that is,
the nonlinear hydrodynamics equations) in favor of 
a somewhat artificial Monte
Carlo dynamics. In particular, some of the results obtained in the
present study indicate that the nonlinear feedback
mechanism that causes the growth of relaxation times in the MC description is
not operative when the artificial Monte Carlo dynamics is used. 
However, since the Monte Carlo method obeys the principle of detailed balance,
it should provide a correct description of the process of thermal activation 
over free energy barriers. Therefore, 
a Monte Carlo calculation of the characteristic time for transition 
from one local minimum of the free energy to another one can
be expected to provide
a correct estimate of the
height of the free energy barrier that separates the two local minima.

 In this work, we use this Monte Carlo dynamics to study the process of
thermally activated transitions from a shallow glassy minimum of the free energy
to regions near
other, more ordered minima. On general grounds, one expects that a system
initially prepared in a glassy state with relatively high free energy would
evolve toward a crystalline state which corresponds to the global minimum of
the free energy at the densities of interest here. It is also expected that 
the transition of 
the system from the initial glassy minimum to a crystalline one would not, in
general, be a single-step process: the system would sample a number,
possibly large, of 
additional free energy minima during its evolution toward equilibrium. This
manifestly non-stationary process of approach to equilibrium is what we are
interested in here. In our study, we start the system off in one of the 
shallow  glassy free energy 
minima obtained from our Langevin work \cite{us3} and use the Monte Carlo
procedure to evolve it in time (as measured in Monte Carlo steps)
until it makes a transition to a different
free energy minimum. Using a set of criteria described in detail in the
next section, we find that the new minima to which the system
moves generally have a lower free energy and are more 
``crystalline''.
The 
characteristic time for such transitions
increases as the density is increased. We find a second ``crossover''
density $n^*_y$ near which the time for transition from one glassy minimum
to another increases very sharply and becomes longer than the times accessible
in our simulation. A limited investigation of the dependence of this barrier
crossing time scale 
on the size of the simulation sample does not show any evidence
for its growth  with increasing system size. As far as we know, this
study is the first one that provides explicit and unambiguous
information about the time scales for transitions between different glassy
free-energy minima of the hard sphere system.
In molecular dynamics (MD)
simulations \cite{WO,WA}, which provide most of the existing
numerical data on the dynamics of this system, it is not possible to
determine unambiguously whether the system fluctuates near the uniform liquid
minimum of the free energy or near an inhomogeneous glassy minimum
at a particular instance of time. 
For this reason, MD simulations do not provide information about the time
scales associated with transitions between different glassy minima.

The remaining part of this paper is organized as follows. Section II contains
descriptions of the Monte Carlo method,
the procedure used by us to monitor the transition of the system from one
local minimum to another, and the criteria we use to specify the degree of
crystalline order present in a free energy minimum. The results obtained 
from this
work are described in detail in section III. In section IV we summarize the 
main
conclusions, compare our findings with MD results and
discuss the implications of our results on the interpretation of experimental
and numerical data on the dynamics of supercooled liquids.

\section{Methods}

The basic ingredient of our study of the dynamics of a dense hard sphere
system is the functional relating the free energy $F$ to the density field
$\rho({\bf r})$. We take for this the Ramakrishnan-Yusouff \cite{RY} (RY)
functional:
\begin{eqnarray}
F[\rho] &=& F_l[\rho_0]+ k_B T \left[ \int{d {\bf r}\{\rho({\bf r})
\ln (\rho({\bf r})/\rho_0)-\delta\rho({\bf r})\} } \right. \nonumber \\
&-& \left. (1/2)\int{d {\bf r} \int {d{\bf r}^\prime
C({|\bf r}-{\bf r^\prime|}) \delta \rho ({\bf r}) \delta
\rho({\bf r}^\prime)}} \right],
\label{RYF}
\end{eqnarray}
where $\delta \rho({\bf r})\equiv \rho({\bf r})-\rho_0$ is the deviation
of the number density from its average value $\rho_0$, $k_B$ is Boltzmann's
constant, $F_l$ the free energy of a uniform liquid of density
$\rho_0$, $T$ the temperature, and $C(|{\bf r}-{\bf r^\prime}|)$ is
the direct pair correlation function \cite{PY} which is taken to be given in the
Percus-Yevick\cite{PY} approximation for hard spheres. It can then
be expressed in dimensionless form in terms of the density parameter
$n^*\equiv \rho_0 \sigma^3$.

In our previous work, \cite{us1,us2,us3,us4} we have studied the dynamical
properties
of this model by introducing the appropriate Langevin dynamics. Briefly,
one introduces the additional current field ${\bf g}({\bf r})$ and the
total free energy:
\begin{equation}
F_{Tot}[\rho,{\bf g}]=(m_0/2) \int{d {\bf r}{|{\bf g}({\bf r})|^2
\over \rho_0}}+F[\rho]
\label{FTOT}
\end{equation}
and then one proceeds to derive the Langevin equations by using the fields
$\rho$ and ${\bf g}$ as the ``slow'' variables, evaluating the needed
Poisson brackets, and adding the appropriate dissipative and noise terms.
This derivation\cite{us4} results in a
system of integrodifferential equations with stochastic terms, which can
then be solved numerically.

The Langevin formulation of the dynamics is appropriate to the physical
situation where deviations of the local density from its spatial average are
small.
As was shown in Ref.(\onlinecite{us3}), the Langevin dynamics used in our
previous studies is inadequate when very large scale
density fluctuations are present.
In practice, this means\cite{us3,us4} that the
Langevin dynamics will lead to equilibration at very long times for densities
$n^*\le n^*_x$, where $n^*_x \simeq 0.95$. At higher densities, if one starts
with liquidlike initial conditions, the Langevin dynamics does not evolve
all the way to equilibrium: after a certain time has elapsed, the density
inhomogeneities grow strong, indicating a transition of the system from the
vicinity of the uniform liquid minimum to the neighborhood of 
one of the inhomogeneous local minima of the free
energy.
As was seen in Ref.(\onlinecite{us3}),
the time $\tau^\prime$ at which this transition 
occurs can be characterized as the
time at which the
quantity $\delta F$ defined as:
\begin{equation}
\delta F \equiv F[\rho]-F_l[\rho_0]
\label{delf}
\end{equation}
becomes negative. This quantity can be evaluated as a function of time
given the density distribution at that time and Eq.(\ref{RYF}).
This has a simple physical interpretation: at higher
densities the system is trying to reach an inhomogeneous state which
may be crystalline or glassy.
The value of the free energy when this inhomogeneous state is reached
will be, at higher densities, lower than the free  energy of the reference
homogeneous liquid. Therefore $\delta F$ will start to become negative as
soon as the larger local density fluctuations characteristic of a 
crystalline or glassy state
become prevalent.

An alternative method to study the dynamics of the system with free energy
(\ref{RYF}) which is valid at times $t>\tau^\prime$ must
be sought. Here we use a
Monte Carlo dynamics implemented in a way that we now describe:

We consider a computational lattice as used for the Langevin
study: a cubic lattice with lattice constant $h$.
The lattice constant of the computational lattice is taken to be
incommensurate with $\sigma$, which has the effect of inhibiting the
formation of an ordered crystal.
In this lattice the density field is described as a set of
$N^3$ numbers $\rho_i$ representing the density at each lattice point
multiplied by $h^3$. We
sweep the lattice considering each site $i$. Given the site $i$ we select
at random a site $j$ which is among the neighbors of $i$, that is, such that
the distance between these sites is less than the hard sphere diameter
$\sigma$. We then evaluate the quantity $s_{ij}=\rho_i+\rho_j$ ,
generate a random number $p$ distributed uniformly between 0 and 1, 
and attempt to change the values of
$\rho_i$ and $\rho_j$ to $p s_{ij}$ and $(1-p)s_{ij}$ respectively.  Clearly
this exchange conserves particle number. The exchange is accepted or rejected
according to a standard Metropolis algorithm with:
\begin{equation}
W_{ij}=\exp-[\lambda \beta \Delta F]
\end{equation}
where $\Delta F$ is the change in the free energy $F$ as given in (\ref{RYF})
which would be produced by the attempted exchange, $\beta$ the usual inverse
temperature, and $\lambda$ the appropriate ratio between the number of lattice
degrees of freedom and the number of spatial degrees of freedom of the actual
continuum system: $\lambda \equiv 1/(3 n^* a^3)$, where $a \equiv h/\sigma$
is the ratio of the  computational lattice
constant to the hard sphere diameter.

In an approximate sense, this 
dynamics may be thought to represent the diffusion processes that would
characterize the later stages of the evolution of the system. It has two
additional advantages: it can handle density fluctuations of any size, and
it is algorithmically very efficient. One inconvenience of the procedure we
follow, however, is that the Langevin and Monte Carlo time scales are not
directly comparable. The difficult
question of their comparison has been discussed
in the past for simpler systems\cite{MS,VMO}. We will return to
this question in section III. For many purposes it suffices to
make comparisons in terms of time scales obtained from the same dynamics
and we take this course when possible.  

As it follows from the Introduction, the present study covers densities
$n^*\ge0.94$. In this density range, one can proceed with
Langevin dynamics until one
reaches the point where $\delta F=0$ as explained above. We then construct
our initial conditions for the Monte Carlo dynamics in the following
way: we consider the state of the system at Langevin time $\tau^\prime$ (the
Langevin simulation having been started\cite{us1,us2} from a perfectly
disordered state), and,
as in Ref. (\onlinecite{us3}), we consider the density configuration of
the system at that time. We use this configuration as the input in a 
minimization routine \cite{cd1} 
that determines the minimum in the free energy $F$
(Eqn.(\ref{RYF})) which is closest to that configuration in the
sense that it is the local minimum the system moves to when one
attempts to find a minimum by making successive small changes in the
density in a way that always lowers the free energy. The free energy
minimization procedure used by us is similar in spirit to the conjugate
gradient method developed by Stillinger and Weber \cite{sw}, but with a crucial
difference. The work of Stillinger and Weber involves finding local
minima of the Hamiltonian of the system under study, whereas our 
procedure finds local minima of the free energy which is obviously the 
correct quantity to consider at non-zero temperatures. The free energy
minimum obtained as the output of the minimization routine corresponds, 
evidently, to a negative value of $\delta F$ when the input is chosen as
described above. This
configuration is then that which corresponds to the local
minimum of the free energy ``basin'' in configuration space that the system
is in when the density fluctuations begin to be large, that is, when 
the system has dynamically evolved away from the uniform liquid minimum. The
configurations at these local minima are \cite{us3} glasslike. We take
these local minimum configurations (obtained at each density studied) as the
initial configurations for the Monte Carlo simulations in the main 
studies reported in this work.
Thus, our initial conditions correspond to having the system in a local
minimum of $F$ which is close to the liquid state. Our objective here is
to find out how long the system remains in the basin corresponding to
such minima as it evolves. We call the time characterizing this evolution
$\tau_1$ and it is one of the main quantities of interest here. To determine
$\tau_1$ we begin our Monte Carlo study with the initial conditions described
above. As we proceed, we monitor the value of the free energy, and other
relevant quantities. At periodic intervals, we run the minimization routine
to find out if it still leads to the initial configuration. When it no
longer does (the precise criteria will be given in the next section), we
know that the system has evolved to another basin of the free 
energy, and that it will eventually evolve to another, more ordered
state after possibly sampling a number of additional free energy
regions.

We emphasize that in the work described here, the time evolution of
the system is simulated by the Monte Carlo dynamics only. The Langevin dynamics
used in our previous studies \cite{us1,us2,us3} is used (in conjunction with
the free energy minimization routine, as described above) 
only in the determination of the initial
configurations to be used in the Monte Carlo simulations. Since our goal in the
present study is to estimate the characteristic time for thermally
activated transitions from shallow glassy minima to deeper, more ordered ones,
a proper choice of  the initial state is important. However, the particular
way in which the initial state is obtained should not matter.
The combination of Langevin
dynamics and free energy minimization described above provides a convenient way
of locating appropriate initial configurations. This procedure, however, is
obviously not unique. In fact, as described in the next section,
we also
consider initial configurations corresponding to similar (negative)
values of $\delta F$ and density distributions to those described above,
but obtained simply by a random search for local minima instead of
through the Langevin intermediate step. We also note that in the simulations 
described
in this paper, the free energy minimization procedure is used only to
determine the minimum near which the system fluctuates at a given time: it
plays no role in the simulation of the time evolution of the system.

The free energy $F$ has a large number of minima at the densities of interest.
The state of the system at such minima is to a varying degree glassy or
crystalline, and it is important to have at least a rough criterion to separate
the more glassy from the more ordered states. The minima can be characterized
first by the value of the free energy. This is conveniently done
for computational (i.e. finite) systems in
terms of the dimensionless free energy per particle $f$:
\begin{equation}
f[\rho]= \beta F[\rho]/(n^* N^3 a^3)
\label{f}
\end{equation}
One expects that smaller (i.e. more negative) 
values of $f$ would correspond, at the same density,
to more ordered states. One would like to verify and quantify this  statement.
The primary detailed measure of order is the spatial correlation function
$u(r)$,
(as defined in Eq.(17) of Ref.(\onlinecite{us3})),
which can be calculated from the density distribution at each
minimum. This function describes correlations of the time-averaged local
densities at two points as a function of their separation. 
In Fig.\ \ref{fig1} we show two examples
(solid and dashed lines) of this correlation function
for two different free energy minima at the same density ($n^*=0.95$). These
were obtained numerically for the same computational lattice ($N=15$ and 
$a=1/4.6$) as was used in our previous Langevin work. One can see from this 
figure that the height of the first maximum is not a very good measure of
the degree of ordering, but that the second and particularly the third
maxima are markedly more  prominent for the minimum with the lower free
energy. While the two minima shown in
these two curves have different degrees of short-range order, neither of them is
truly crystalline. To make this point clear, we have shown in Fig.\ \ref{fig1}
(symbols and dotted line)
the function $u(r)$ for a perfectly crystalline minimum 
obtained for a sample with
$N$ = 12 , $a$ = 0.25, and the same density. (We could not find any exactly
crystalline minimum for
samples with $N$ = 15 and $a$ = 1/4.6, 
due to the incommensurability
\cite{us1,us2} of the computational lattice with a fcc one). The function $u(r)$
for the crystalline minimum has very sharp peaks at values of $r$ corresponding
to the lattice vectors of a fcc structure and is very close to zero at all other
values of $r$. These features are clearly
very different from what is found for the
other two minima.

One can proceed with this study in a more systematic way. In Fig.\ \ref{fig2}
we show, for the same density and computational lattice as in the previous
Figure, the relation between the value of the free energy at a given
minimum and three different measures of the degree of order at that
minimum. The comparison includes nine different minima. The quantities
plotted are the height of the second peak in $u(r)$ (which is one tenth
of the quantity represented by the top line in the Figure), the increase in
the average density which occurs as one evolves (under the minimization
procedure) from the homogeneous liquid state to the minimum (middle line), and
the height of the first peak (which is half of the value plotted as
the lower line). We can see that, while the relationship is not simple, it
is generally true that lower free energy values are associated with
a higher degree of order. In particular, the free energy minimum of 
Fig.\ \ref{fig1} with $f$ = -2.12 corresponds to a state at the left end of
Fig.\ \ref{fig2}, whereas the one with $f$ = -1.78 sits at the right end.
In this work, when we refer to glassy free energy minima 
corresponding to ``more ordered'' or ``more disordered'' states we mean
states near the left and right ends, respectively, of Figure\ \ref{fig2}
or of the corresponding situations at other densities. A similar 
classification of glassy states on the basis of the degree of short-range
positional order was also made in MD studies \cite{WO} where
it was found that glassy states obtained after a fast quench from high
temperatures or low densities generally exhibit a lower degree of positional
order than those obtained after a slow quench. The differences in the structures
of glassy minima corresponding to the two ends of Fig.\ \ref{fig2} are, in fact,
quite similar to the differences in the structures of the ``slow'' and ``fast''
glasses of Ref.(\onlinecite{WO}).
As explained above,
our Monte Carlo
runs use one of the more disordered states (typically those obtained
from Langevin dynamics and the procedure explained above) as the initial condition.

We turn now to the results in the next section.

\section{Results}

In this section we present the results obtained from the Monte Carlo 
simulations.
One of our main objectives is to find out whether our hard sphere system,
if initially quenched in a free energy minimum corresponding to a glassy
state, will evolve towards a more ordered final state.

Except for the investigation of size effects described later in this
section, our Monte Carlo work was done for the same system as the Langevin
work of Refs.\ \onlinecite{us1,us2,us3,us4}. We use the same lattice size ($N=15$),
the same incommensurate value of $a$, and, in all $N=15$ cases, an initial state
obtained from the same Langevin runs using the procedure indicated in the
previous section.

We focus on the quantity $\tau_1$. As explained in the previous section,
this quantity measures the time the system spends in an initial, shallow,
``glassy'' minimum. To define $\tau_1$ and to evaluate it explicitly we follow
the following algorithm: We begin the MC simulation, with the initial
conditions as described, and at periodic intervals $t_i$ separated by
intervals $\Delta t \equiv t_{i+1}-t_i$ (all
times from now on are given in terms of Monte Carlo attempts per site,
MCS) we save the current density configuration and run the minimization
routine to find out if the system is still within the initial basin.
We wish $\tau_1$ to represent the time at which the system definitely
leaves the initial basin, to move to other regions of phase space on
its way, eventually, to a more ordered state (presumably after sampling
a large number of minimum basins in phase space). To prevent fluctuations,
arising from situations where the system is evolving along saddle type
regions separating minima, from giving incorrect small values of $\tau_1$
for a given run, we define $\tau_1$ algorithmically as the time after
which the system is found by the minimization procedure at a minimum
different from the initial one for $N_i=3$ consecutive times $t_i$ with
$\Delta t =400$ MCS. By continuing the runs to longer times, we find that
with this criterion the system does not return to the basin of attraction
corresponding to the initial minimum. Such continuing runs were used to
determine by trial and error the appropriate values of $\Delta t$
and of $N_i$. In addition to thus representing the
desired physical quantity, this procedure has the advantage of reducing
run to run fluctuations.

The results for $\tau_1$ obtained by the above procedure, and for the
system and lattice sizes described, are plotted in Fig.\  \ref{fig3}.
For $n^* \le 0.98$ the symbols plotted represent the average over
five runs, and the error bars the standard deviation. For $n^*=0.99$
only one run out of five led to a finite $\tau_1$, as defined above, with
a value corresponding to the lower end of the
arrow like ``error bar'' plotted. The other
runs yielded only lower bounds $\tau_1 >$ 120,000 MCS, that is, the
system remained in the domain of attraction of the initial minimum. We did not
need to calculate $\tau_1$ for $n^* \le 0.93$ because our previous studies
\cite{us1,us2} showed that for such densities, a system initially prepared
in the liquid state, remains in the vicinity of the uniform liquid minimum
for all time scales accessible in our Langevin simulations.
We see from Fig.\ \ref{fig3} that $\tau_1$ increases slowly
for $0.94 \le n^* \le 0.98$, but it then grows extremely fast and at
the vicinity of $n^*=0.99$ it becomes longer than any reasonable measurement
time. Thus, there will be a density $n^*_y \approx 0.99$  such that, if
the system is quenched to a density $n^* > n^*_y$ then it cannot reach
the equilibrium crystalline state in time scales accessible in simulations.
It would be interesting to determine whether the observed growth of
$\tau_1$ with increasing density is described by one of the various
functional forms (e.g. Batchinski-Hildebrand \cite{BH} form,
Vogel-Fulcher \cite{VF} form and power-law form) used in
the literature to describe the growth of time scales in dense
liquids. Due to the limited accuracy and range of our data for
$\tau_1$, we did not attempt to fit them to these functional forms. Such 
fits would be meaningful only if much more accurate data for $\tau_1$ 
spanning several decades of its growth were available.
Given the computational complexity of the numerical method used here
(approximately 200 CPU hours of a Cray 2
machine were required for carrying out the calculations described here), 
and the very high degree of statistical accuracy required to distinguish
among the proposed forms,
generation of such data appears to be computationally
impossible at the present time.

As we monitor the system to evaluate $\tau_1$, we also keep track of other
relevant quantities. In particular, we continued many of the runs to times 
well beyond $\tau_1$ to verify the condition of nonreturn to the
vicinity of the initial state. We found that over time 
scales of the order of several times the value of $\tau_1$, the system samples 
several free energy minima (more accurately, the basins associated
with several free energy minima). The characteristic time spent in one of 
these basins shows a wide variation, with values ranging from a small fraction
of $\tau_1$ to several times the value of $\tau_1$.  
The results seem to hint to the existence of a continuum or a hierarchy
of time scales, rather than a single scale characterized by some dwelling
time.

We monitored also the free energy $F$ as a function of time.  The results
are very suggestive. We show here the results for four runs, two at
$n^*=0.94$ (Fig.\ \ref{fig4}) and two at $n^*=0.95$, (Fig.\ \ref{fig5}).
We see that the free energy first increases, and eventually begins a
slow decrease, passing therefore through a rather shallow but definite
maximum. We find that the time at which this maximum occurs corresponds
to a very good approximation to the value of $\tau_1$ obtained for the
particular run under consideration. In other words: at times earlier
than that at which the maximum in $F$ occurs, the minimization
routine indicates that the system is still in the phase space neighborhood
of the initial minimum, while for times beyond the maximum, it is
elsewhere in phase space. This meshes well with the intuitive
observation that the system must cross over some free energy barrier
in order to move from basin to basin in phase space. The lack of clearly
defined maxima at latter times in these two  plots agrees well with the
remark in the last paragraph that there is no statistically well defined
single time scale for the dwelling time in different basins beyond the
original one.

We have also monitored the correlation function $u(r)$ at all the different
minima reached by the sytem in any given run. With a few exceptions, the new
minima to which the system moves under the Monte Carlo dynamics are found to
have lower free energy and a higher degree of order than the minimum from 
which the run was started. All the minima corresponding to points 
near the left end of Fig.\ \ref{fig2}
were obtained in this way. Although the minima near which the system 
fluctuates for times greater than $\tau_1$ usually have free energies lower 
than  that of the initial minimum, the measured free energies plotted in
Figs.\ \ref{fig4} and \ref{fig5}
for values of $t > \tau_1$ are higher than the free energy at $t$ = 0.
This result, which at first may seem paradoxical,
may be understood in the following way.
The state of the system at time $t$ = 0 coincides with the configuration at a
free energy minimum. So, the free energy at $t$ = 0 shown in Figs.\
 \ref{fig4} and \ref{fig5}  is
just the free energy of the minimum from which the Monte Carlo run is started. 
At later times, the Monte Carlo procedure generates fluctuations about the
free energy minimum. Therefore, the free energy measured at a later time 
may be viewed as the
sum of the free energy of the minimum near which the system happens to reside
at that time and the free energy associated with fluctuations about this 
minimum. The second contribution, which is always positive, may be roughly
estimated to be about $1.5 k_B T$ per particle in a harmonic phonon-like
approxamation. Due to the presence of this additional contribution, the measured
free energy at times greater than $\tau_1$ remains higher than the free energy
at $t$ = 0 unless the free energy of the minimum near which the system
fluctuates at these later times is very much lower than that of the initial
minimum. This may happen at very long times when the system reaches the
vicinity of the crystalline state. The time
scales accessible in our simulations are, however, too short for this behavior
to be seen. In our simulations, we can study only the initial part of the
evolution of the system toward the crystalline state which is the
true equilibrium state of the system at the densities considered here. 

We turn now the question of the dependence of $\tau_1$ on the size of the
simulated sample. It is important to check
whether $\tau_1$ is size independent or not. If, for example, it was found
that $\tau_1$ increases with system size, that is, if it were divergent in the
thermodynamic limit, then we would have to conclude that in the
region of interest the crystalline state is never accessible. This is
not the case. We have computed $\tau_1$ for different sample sizes 
at the density $n^*=0.96$, a
value which is in the intermediate range  of densities considered. Since
the Langevin results are available only for samples with $N$ = 15 
and it would have been
too expensive to obtain the necessary results for other values of
$N$, we used for our initial conditions for $N \ne $ 15 
the density distribution at
shallow free energy minima (as explained in connection with
Fig.\ \ref{fig2}) obtained from a large number of minima
located using the minimzation routine and random initial
conditions. It was not possible to obtain shallow minima with
glassy density distributions for the same value of $a$ used at $N=15$
because the commensurability conditions are different for different values
of $N$. However, the
results quoted here for $N=12$ and $N=18$  do correspond to the
same value of $a$ (=0.25) and can therefore be directly compared with each
other. The correlation functions for the minima actually
used are shown in  Fig.\ \ref{fig6}. One can see that despite the 
difference in size the distributions are extremely similar. 
We again averaged the results for $\tau_1$
over five runs and we found that this quantity is, within statistical error,
independent of size, even though the number of particles is $3.375$
times greater for the larger value of $N$. We conclude then that at least for
$n^* < n^*_y$, $\tau_1$ remains finite in the thermodynamic limit. This result
is in agreement with suggestions made in earlier studies \cite{pgw,HW} of 
similar systems.

Considering the results described above and taking also into
accouunt those obtained from our previous
work \cite{us1,us2,us3,us4}, we obtain the following description of the
dynamics of a dense hard sphere system in a rather wide density
range. For the discretized version of the 
Ramakrishnan-Yussouff free energy functional considered by us, the equilibrium 
transition to a crystalline phase occurs at a density $n^*_f \simeq$ 0.83
\cite{cd1,cd2}. The uniform liquid minimum of the free energy, however, remains
locally stable for a large range of densities above $n^*_f$. Our Langevin
work \cite{us1,us2}
shows that a system initially prepared in the liquid state remains in the
direct vicinity of the liquid minimum over all accessible time scales if the
density is lower than a {\it first crossover density} $n^*_x \simeq $ 0.95.
The dynamic behavior for $n^* < n^*_x$ is, therefore, governed by small 
fluctuations about the uniform liquid minimum. The results of our
previous study \cite{us1,us2} show that the dynamics in this regime is 
described fairly well by MC theories. For $n^* > n^*_x$, on the other hand,
the system moves 
away from the liquid minimum over numerically accessible time 
scales \cite{us3}. The minima to which the system moves are usually found to
be glassy according to the criteria described in section II. 
At subsequent times, the system  
moves toward states with a higher
degree of
crystalline order because such states generally have lower free energy.
This process is slow because it involves thermally activated
transitions over free energy barriers.
Our present calculation shows that the typical time scale $\tau_1$
associated with this process increases with density. The initial increase
of $\tau_1$ with $n^*$ is slow, but it increases very sharply as $n^*$
approaches a {\it second crossover density} $n^*_y \simeq $ 0.99. At higher
densities, the time scale $\tau_1$ becomes longer than the longest time scales
accessible in our simulation. This observation implies that a hard sphere liquid
quenched from a low density to a density $n^* > n^*_y$ would remain stuck in
a glassy free energy minima for all time scales accessible in numerical 
simulations. For $n^* < n^*_y$, on the other hand, the system would move to
states with a higher degree of crystalline order 
during simulational time scales. Our calculation 
indicates that the time scale $\tau_1$ is an intrinsic property of the
system in the sense
that it does not depend strongly on the size of the system
considered in the simulation. Therefore, the general description outlined
above should remain valid in the thermodynamic limit.

We also carried out a 
limited study of the Monte Carlo dynamics of the system
in the vicinity of the uniform liquid minimum for a number of
densities in the range $0.93 \le n^* \le 0.99$. In these simulations,
the initial state of the system was taken to be uniform. 
The Monte Carlo procedure was then used to
simulate the dynamic behavior for a few hundred MCS. The relaxation
time obtained from the observed initial decay of the
nonequilibrium density autocorrelation function from the uniform state
(defined as $<(\rho_i(t_0+t)-\rho_0 h^3)(\rho_i(t_0) - \rho_0 h^3)>$,
 where
the average is over all sites $i$ and a range of initial $t_0$) 
was found to be very short (a few MCS) and only weakly
dependent of the density for values of $n^*$ in the range studied. 
Thus, the nonlinear feedback
mechanism of MC theories which predicts a rapid growth of the relaxation
time of small-amplitude density fluctuations near the liquid minimum
does not appear to be operative for the Monte Carlo dynamics.
An explanation of this result, which may appear
surprising, is provided in the next section. Since the dynamic behavior of
the system in the vicinity of the liquid minimum  was studied in
detail in our previous work \cite{us1,us2,us3,us4} using a dynamics
(Langevin equations of fluctuating hydrodynamics) which is physically
appropriate for this liquid-like situation, we did not think it useful to
carry out
a more detailed study of the Monte Carlo dynamics, which is clearly not
the appropriate one for describing the physics in this regime.                

We now address the issue of relating the Monte Carlo time unit (1 MCS) to
the Langevin time unit, which is close to \cite{us1} the Enskog
collision time. It would appear that a ``calibration'' of
the Monte Carlo time step might be obtained by comparing the results
obtained for the same physical time scale by using both Monte Carlo and
Langevin dynamics. As mentioned in the Introduction,
the Langevin scheme can not be used to calculate the time scale $\tau_1$.
However, the relaxation time of small-amplitude density fluctuations
near the liquid minimum can be calculated for both Monte Carlo and
Langevin dynamics, and we have the results for the Langevin
relaxation time for $n^* \le 0.93$ from our previous work
\cite{us1,us2} and for the Monte Carlo relaxation time from the
present work. For example, the Langevin relaxation time at $n^* =
0.93$ is 2300 in the units of Ref (\onlinecite{us1}), whereas the
Monte Carlo relaxation time at the same density is about 2 MCS. These
results may be used to obtain a rough correspondence between the two
time units. However, such a comparison must be viewed with extreme
caution because one should not really compare nonequilibrium time
scales obtained from two completely different dynamics. The 
difficulty in making such a correspondence is exemplified by the 
fact that the correspondence factor
obtained this way would depend strongly on the 
density because the Langevin relaxation time increases rapidly with
increasing density while the Monte Carlo relaxation time remains 
nearly constant in the density range studied.

A discussion of the relative merits of the Monte Carlo method used
here and the standard MD method would be interesting from a
methodological point of view. It is difficult to determine
which algorithm is ``faster'' because
the answer seems to depend on the nonequilibrium process being studied. 
The process of crystallization
appears to be faster in MD than in our Monte Carlo simulations.
MD simulations often show \cite{WO}
nucleation of the crystalline state during simulation time scales at
high densities, whereas we seldom see crystallization in our Monte Carlo
simulations at similar densities. On the other hand, small
fluctuations near the
uniform liquid minimum relax faster in the Monte Carlo
dynamics than in MD. 
Our Monte Carlo method has the
advantage of being able to relate the observed behavior to different
local minima of the free energy. This can not be done in MD
simulations.

\section{Conclusions and Discussion}

The main result obtained in this work is the determination of the time scale
$\tau_1$ associated with activated transitions between glassy metastable
minima of the free energy of a dense hard sphere system. The observed 
dependence of
$\tau_1$ on the dimensionless density parameter $n^*$ (Fig.\
\ref{fig3}) establishes the
existence of a crossover density $n^*_y \simeq 0.99$ with the property that
systems quenched to densities higher than $n^*_y$ would remain stuck in a
glassy state for all time scales accessible in simulations. We emphasize that
the growth of $\tau_1$ with density is {\em not} caused by the MC mechanism
which is responsible for 
the growth of relaxation times for values of $n^*$ lower than
the other crossover density $n^*_x$ mentioned above. That the
nonlinear feedback mechanism of MC theories is not present in the Monte
Carlo dynamics used in the present work is explicitly demonstrated by the
observation that
the relaxation time of small-amplitude density fluctuations near the
liquid minimum is very short (of the order of a few MCS) and approximately 
independent of the density  when the system  
evolves under the Monte Carlo dynamics. This result arises because in
in the Monte Carlo algorithm we use, two cells which are within a
distance $\sigma$, but not necessarily nearest neighbors are selected
at random and the densities at these two cells are changed by random
amounts. These changes, although conserving the total mass, are not
governed by an equation of continuity containing the divergence of a
current field. Therefore, the physics that arises as a consequence of
the coupling of the density field to the current field through
the equation of continuity is not
expected to be present in the results obtained by using this algorithm.
Since the nonlinearities which lead to the feedback mechanism of MC
theories come from the coupling of the density field to the current
field and from the nonlinearities present in the equation of motion
of the current field, it is not surprising that the slowing down of
the kinetics predicted in MC theories (and observed in our previous
Langevin work \cite{us1,us2}) is not found in the Monte Carlo dynamics 
used in the present work.

Therefore, the present work establishes the
existence of two distinct regimes for the dynamics of the dense hard sphere
liquid: in the first regime, $n^* \le n^*_x$, which covers the first few orders
of magnitude of the growth of relaxation times, the slowing down of time
scales is a consequence of the nonlinear feedback mechanism described by MC
theories. In the second regime, $n^* > n^*_x$, the slow relaxation arises from
thermal activation over free energy barriers between different inhomogeneous
minima of the free energy. The growth of relaxation times with increasing 
density in this regime must
be attributed to an increase of the characteristic height of these free energy
barriers.

The results of the present study are in qualitative agreement with the 
observations of existing MD studies \cite{WO,WA} of the same
system. MD
simulations show that the system spontaneously
freezes into a near-crystalline state if it is allowed to  evolve in time at
densities higher than a ``first critical density'' $n^*_s \simeq$ 1.08. If, on
the other hand, the system is rapidly quenched from the liquid state at a 
density lower than $n^*_s$ to a density above a ``second critical density''
$n^*_t \simeq$ 1.2 (which is close to the random close packing density, $n^*
\simeq$ 1.23), then it ends up in a glassy state with little crystalline order.
In our previous work \cite{us3}, we identified the density $n^*_x$ with $n^*_s$.
The present work suggests that the new crossover density $n^*_y$ should be 
identified with 
the density $n^*_t$ found in MD simulations. As mentioned
in section III, we find that the minima to which the system moves over the
time scale $\tau_1$ are generally more crystalline than the initial ones.
This result is consistent with the observation \cite{WO}, made in MD
simulations, that the degree of order present in the glassy states
obtained by quenching the system to densities higher than $n^*_t$ increases
as the quenching rate is decreased. The value of
$n^*_y$ obtained here ($n^*_y \simeq$ 0.99) 
is substantially lower than the result for $n^*_t$
($\simeq$ 1.2) obtained in MD simulations. As mentioned in
Ref(\onlinecite{us3}), this difference is likely due to the fact that the
value of $n^*$ at which the discretized version of the free energy functional
used by us exhibits a thermodynamic crystallization transition ($n^*_f \simeq $
0.83) is substantially lower than the crystallization density obtained in
MD simulations ($n^*_f \simeq$ 0.943)\cite{WO}. Indeed, the
ratios $n^*_x/n^*_f$ and $n^*_y/n^*_f$ that we find for our system are
quiet similar to the corresponding ratios $n^*_s/n^*_f$ and $n^*_t/n^*_f$
as found in the MD work.

We conclude with a few remarks on the implications of the results of the present
study on the interpretation of experimental data on the dynamics
of good glass forming systems. In contrast to the hard sphere system 
for which near-crystalline states are relatively easy to reach 
if $n^*$ lies between $n^*_x$ and $n^*_y$ (as indicated
by MD results and our observation that the minima to which the
system moves during its evolution under the Monte Carlo dynamics usually have
higher degrees of crystallinity), the ordered state is expected to be highly
inaccessible in good glass forming liquids such as two (or more) component
mixtures. In such systems, the transitions described by the time scale $\tau_1$
may not take the system closer to the ordered state. Instead, the system may 
continue to wander among various glassy minima of the free energy. In the long
time limit, such a system may still be a liquid in the sense that the 
time-averaged local density may still be 
uniform. However, the dynamics of such a 
liquid (governed by the time scale $\tau_1$ defined above) 
would be quite different from that of the same system at densities lower than
$n^*_x$ , for which the dynamics is governed by the MC time scales. We suggest
that this picture may provide an explanation of the ``crossover'' in the
dynamics observed
\cite{jack,angell} near the ideal glass transition temperature $T_c$ 
in experiments on fragile liquids.

A question that naturally arises in this 
context is whether the present study has anything to say about the possibility
of a true second order glass transition in such systems. Such a transition
would correspond to a true divergence of the time scale $\tau_1$ in the
thermodynamic limit. Our investigation of the dependence of $\tau_1$
on sample size does not show any evidence for such a divergence. However, our
study of sample-size dependence was rather limited in scope. In particular,
we only considered glassy local minima which are rather shallow. We found some
evidence (although this aspect was not studied in any detail) which suggests
that the time scales for transitions from deeper glassy minima are longer than
the values of $\tau_1$ quoted above. A more comprehensive study of this and
other related issues would be most interesting.

\acknowledgments

One of us (CD) is grateful to the Department of Physics, University of
Minnesota and Minnesota Supercomputer Institute for support and hospitality.

\begin{figure}
\caption{The radial correlation function (called $u(r)$ in the text) as
a function of distance $r$ (measured in units of $\sigma$). All
curves
correspond to the same density $n^*=0.95$. The solid and
dashed lines (without symbols) correspond to glassy
states with $N=15$ and $a=1/4.6$. The solid line refers to a
minimum having a free energy per particle $f=-1.78$, while the dashed
line is for the case of a lower $f$, $f=-2.12$. For comparison we include
also a plot (symbols and dotted line) corresponding to the crystalline
case with $N=12$ and $a=0.25$.}
\label{fig1}
\end{figure}
\begin{figure}
\caption{The correlation between three measures of degree of order and
the free energy per particle at a given minimum. This plot is for $n^*=0.95$,
similar plots can be made at other densities. Three different quantities are
plotted as functions of
the free energy per particle $f$. These quantities are:
Top (diamonds and solid line), the height of the second peak in $u(r)$
times ten. Middle, (squares and short dashes): percentage density increase
(see text). Bottom (crosses and long dashes): height of the first peak
in $u(r)$ times two. The factors are chose so as to clearly separate the
three plots. The symbols are in all cases the actual results for nine different
free energy minima and the lines just connect the dots.}
\label{fig2}
\end{figure}
\begin{figure}
\caption{The quantity $\tau_1$ as specified in the text, plotted as a function
of density. At $n^*=0.99$ the lower end of the arrow  represents a lower
bound}
\label{fig3}
\end{figure}
\begin{figure}
\caption{The dimensionless free energy $\beta F$ plotted vs time 
for two runs at $n^*=0.94$.}
\label{fig4}
\end{figure}
\begin{figure}
\caption{The dimensionless free energy $\beta F$ as
in Fig. 4, but for $n^*=0.95$.}
\label{fig5}
\end{figure}
\begin{figure}
\caption{The radial correlation functions $u(r)$ for the free energy minima used
as initial conditions in the Monte Carlo simulations at sizes
$N=12$ (solid line) and $N=18$ (dashed line). The density is
$n^*=0.96$ in both cases.}
\label{fig6}
\end{figure}

%
%

\end{document}